\documentclass[12pt]{iopart}
\usepackage{graphicx}
\usepackage{color}
\begin{document}

\title[Two-photon resonant excitation of interatomic coulombic decay in neon dimers]{Two-photon resonant excitation of interatomic coulombic decay in neon dimers}
\author{A~Dubrouil$^{1}$, M~Reduzzi$^{1,2}$, M~Devetta$^{2}$, C~Feng$^{1}$, J~Hummert$^{1}$, P~Finetti$^{3}$, O~Plekan$^{3}$, C~Grazioli$^{3,4}$, M~Di~Fraia$^{5}$,  V~Lyamayev$^{6}$, A~La~Forge$^{6}$, R~Katzy$^{6}$, F~Stienkemeier$^{6}$, Y~Ovcharenko$^{7}$, M~Coreno$^{8}$, N~Berrah$^{9}$, K~Motomura$^{10}$, S~Mondal$^{10}$, K~Ueda$^{10}$,\\ K~C~Prince$^{3}$, C~Callegari$^{3}$, A~I~Kuleff$^{11}$, Ph~V~Demekhin$^{12}$, G~Sansone$^{1,2}$}
\address{$^1$Dipartimento di Fisica, Politecnico Piazza Leonardo da Vinci 32, 20133 Milano, Italy}
\address{$^2$CNR-IFN, Politecnico Piazza Leonardo da Vinci 32, 20133 Milano, Italy}
\address{$^3$Elettra-Sincrotrone Trieste in Area Science Park, 34149 Trieste, Italy}
\address{$^4$Department of Chemical and Pharmaceutical Sciences, University of Trieste, Via L. Giorgieri 1, I-34127 Trieste, Italy}
\address{$^5$Department of Physics, University of Trieste, Trieste, 34127, Italy}
\address{$^6$Physikalisches Institut Universit\"{a}t Freiburg Hermann-Herder-Str. 3, D-79104 Freiburg, Germany}
\address{$^7$Institut f\"{u}r Optik und Atomare Physik, Technische Universit\"{a}t Berlin, Germany}
\address{$^8$CNR Istituto di Metodologie Inorganiche e dei Plasmi, Area della Ricerca\\ di Roma~1, Monterotondo Scalo, Italy}
\address{$^9$Physics Department, University of Connecticut, Storrs, CT 06268, USA}
\address{$^{10}$Institute of Multidisciplinary Research for Advanced Materials, Tohoku University, Sendai 980-3204, Japan}
\address{$^{11}$Theoretical Chemistry, Institute of Physical Chemistry, University of Heidelberg,\\ Im Neuenheimer Feld 229, 69120 Heidelberg, Germany}
\address{$^{12}$Institut f\"{u}r Physik und CINSaT, Universit\"{a}t Kassel, Heinrich-Plett-Str. 40, D-34132 Kassel, Germany}
\ead{\mailto{demekhin@physik.uni-kassel.de; giuseppe.sansone@polimi.it}}

\begin{abstract}
The recent availability of intense and ultrashort extreme ultraviolet sources opens the possibility to investigate ultrafast electronic relaxation processes in  matter in an unprecedented regime. In this work we report on the observation of two-photon excitation of interatomic Coulombic decay (ICD) in neon dimers using the tunable intense pulses delivered by the free electron laser FERMI@Elettra. The unique characteristics of FERMI (narrow bandwidth, spectral stability, and tunability) allow one to resonantly excite specific ionization pathways and to observe a clear signature of the ICD mechanism in the ratio of the ion yield created by Coulomb explosion. The present experimental results are explained by \emph{ab initio} electronic structure and nuclear dynamics calculations.
\end{abstract}

\maketitle

\section{Introduction}
\label{sec:intro}
Nonlinear interactions between electromagnetic fields and matter are at the heart of several technologies and experimental techniques. Since the first experimental demonstration of second harmonic generation~\cite{PRL-Franken-1961}, increasing laser peak intensity has led to the demonstration of nonlinearities in the laser-matter interaction in the electromagnetic spectrum extending from the THz~\cite{NATPHOT-Schuber-2014} up to the X-ray spectral range~\cite{PRL-Doumy-2011}. In the extreme ultraviolet (XUV) and X-ray spectral range the observation of nonlinear processes has challenged experimental demonstration due to the lack of intense, coherent sources. With the advent of Free Electron Lasers (FELs) operating in the XUV~\cite{NATPHOT-Ackermann-2007} and X-ray spectral range~\cite{NATPHOT-Emma-2010}, a new regime for the investigation of laser-matter dynamics under extreme conditions has become accessible. The high intensity combined with ultrashort pulse duration, approaching the attosecond domain~\cite{NATPHOT-Helml-2014}, makes these sources ideal for the time-resolved investigation of inner-valence and core-shell electron dynamics.

Electron-correlation-driven processes (i.e., processes driven by the interaction between electrons beyond the one-particle approximation) play a fundamental role in the structure and dynamics of atoms, molecules and solids~\cite{Chemphyschem-Sansone-2012}. These are, for example, all types of electronic decay processes of electronically excited states which typically proceed on the few femtoseconds or even subfemtosecond timescale~\cite{NATURE-Drescher-2002}. An important class of such electron relaxation processes is the interatomic Coulombic decay (ICD) which has attracted an increasing interest since its prediction~\cite{PRL-Cederbaum-1997} and first experimental observations~\cite{PRL-Marburger-2003, PRL-Jahnke-2004}.

The ICD phenomenon represents an efficient energy transfer from an electronically excited system to its environment, which uses the energy to eject an electron. The process is driven by the electron correlation and takes place in a variety of weakly bound systems, like van der Waals and hydrogen bonded clusters~\cite{RevThe,RevExp,RevJahnke}, due to its high efficiency. ICD has been observed in large rare-gas clusters~\cite{PRL-Ohrwall-2004}, water dimers~\cite{NATPHYS-Jahnke-2010}, water clusters~\cite{NATPHYS-Mucke-2010}, and even in quantum wells~\cite{PRB-Cherkes-2011} and endohedral fullerens~\cite{PRL-Averbukh-2006}. The occurrence of ICD demonstrates the importance of the chemical environment in electron-driven relaxation process. Moreover, the process represents an efficient source of low energy electrons~\cite{rICDthe,rICDexp} in aqueous environments which play an important role in the mechanisms associated with radiation-induced damage of biological molecules. The investigation of energy relaxation processes, like ICD, will strongly benefit from the possibility to trigger the relaxation dynamics in a selective and efficient way.

\begin{figure}
\centering\includegraphics[scale=0.75]{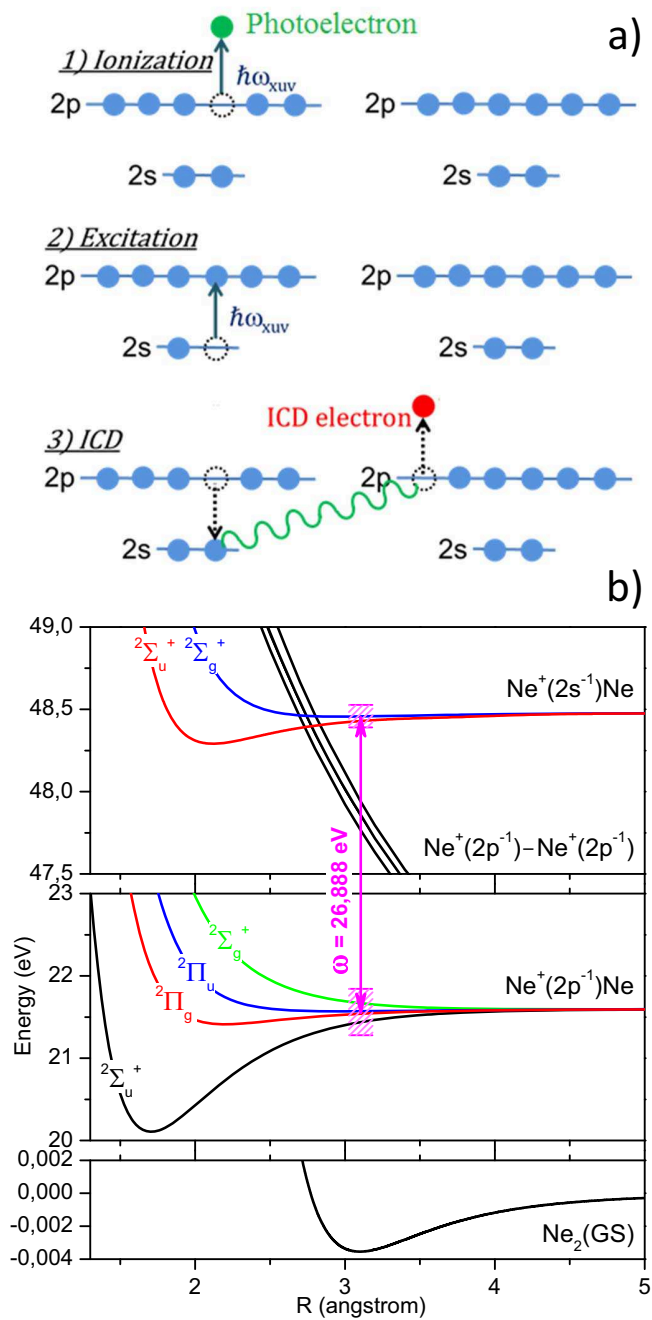}
\caption{a): Schematic representation of the two-photon excitation of ICD. In the \emph{Ionization} step, the $2p$-electron is ionized by a first photon from the XUV pulse with the emission of a photoelectron and population of the OV-ionized $\mathrm{Ne^{+}_2}(2p^{-1})$ states of the dimer. At the \emph{Excitation} step, a second photon is resonantly absorbed leading to the excitation of the IV-ionized $\mathrm{Ne^{+}_2}(2s^{-1})$ states. During the \emph{ICD} step, the $2s$-hole of the initially ionized atom is filled by a $2p$-electron, and the excess energy is transferred to the neighboring atom, leading to the emission of the ICD electron. As a result, the two-site OVOV doubly-ionized dimer $\mathrm{Ne^{+}}(2p^{-1})\mathrm{Ne^{+}}(2p^{-1})$ undergoes Coulomb explosion. b): Potential energy curves of the relevant  neutral ground state, the OV and IV singly-ionized states, as well as OVOV doubly-ionized states of the neon dimer. Fig.~\ref{Fig1}b is adapted from Ref.~\cite{PRL-Demekhin-2011}.}
\label{Fig1}
\end{figure}

Recently, the two-photon excitation of ICD in neon dimers was proposed theoretically~\cite{PRL-Demekhin-2011}. The suggested multiphoton absorption scheme should provide a more efficient triggering of the ICD  with respect to  traditional one-photon ionization schemes. The excitation proceeds through the absorption of two XUV photons as shown schematically in Fig.~\ref{Fig1}a. The first photon ionizes the neon dimer creating an outer-valence (OV)  singly-ionized state (\emph{Ionization} step in Fig.~\ref{Fig1}a), and transferring the nuclear wave packet from the ground state of the neon dimer to one of the cationic states converging at large internuclear distances to the electronic configuration $\mathrm{Ne^{+}}(2p^{-1})\mathrm{Ne}$.  The uppermost singly-ionized state $^2\Sigma_g^+$ is repulsive, leading to the formation of $\mathrm{Ne}$ and $\mathrm{Ne^+}$, whereas the other three potential energy curves  $^2\Sigma_u^+, ^2\Pi_u$, and $^2\Pi_g $ are bound and do not lead to  dissociation (see Fig.~\ref{Fig1}b). The subsequent interaction with the XUV field can lead to the excitation of a $2s$-electron  of the initially ionized atom to the unoccupied $2p$-level (\emph{Excitation} step in Fig.~\ref{Fig1}a). In terms of molecular states, the absorption of a second XUV photon results in the resonant transitions $^2\Sigma_u^+, ^2\Pi_u  \to ^2\Sigma_g^+$ and $^2\Sigma_g^+, ^2\Pi_g \to ^2\Sigma_u^+$. These transitions populate the $^2\Sigma_u^+, ^2\Sigma_g^+$ inner-valence (IV) singly-ionized states of the dimer, which converge at large internuclear distances to the  electronic configuration $\mathrm{Ne^{+}}(2s^{-1})\mathrm{Ne}$  (see Fig.~\ref{Fig1}b). Starting from this state, ICD  can occur (\emph{ICD} step in Fig.~\ref{Fig1}a) leading to the emission of a second electron, population of the two-site OVOV doubly-ionized states $\mathrm{Ne^{+}}(2p^{-1})\mathrm{Ne^{+}}(2p^{-1})$ (repulsive curves in Fig.~\ref{Fig1}b), and, finally, Coulomb explosion of the doubly-charged neon dimers. The ionization, excitation, and relaxation mechanisms can be summarized as follows:
\begin{eqnarray}
Ionization:~&\mathrm{Ne_{2}}
+\hbar\omega & \to \mathrm{Ne^{+}_2}(2p^{-1})+e^-_{PE}\nonumber\\
Excitation:~&\mathrm{Ne^{+}_2}(2p^{-1})+{\hbar\omega}&\to\mathrm{Ne^{+}_2}(2s^{-1})\nonumber\\
ICD:~&\mathrm{Ne^{+}_2}(2s^{-1})&\to \mathrm{Ne^{+}+Ne^{+}}+e^-_{\mathrm{ICD}}.
\label{Eq1}
\end{eqnarray}

As discussed in Ref.~\cite{PRL-Demekhin-2011}, the signature of the ICD in the third step of process~(\ref{Eq1}) is prominently manifested in the spectrum of emitted electrons. It was shown that at XUV intensities below $10^{12}~\mathrm{W/cm^2}$, the electron signal around 1~eV is dominated by the ICD mechanism, while other competing mechanisms, such as two-site two-photon double ionization of the neon dimers, starts playing a relevant role only for higher intensities~\cite{PRL-Demekhin-2011}. In this work, we focus  on the effect of the ICD process on the generation of the singly-ionized neon atoms and on the residual population of the singly-ionized neon dimers. In particular, we observe a clear signature of the excitation of the ICD process in the ratio of $\mathrm{Ne}_2^+$  vs  $\mathrm{Ne}^+$ cations.

The manuscript is organized as follows. In Section~\ref{theor}, we discuss the theoretical approach used to calculate the population of the singly-ionized neon dimers under the excitation by an intense tunable extreme ultraviolet pulse. In Section~\ref{experiment}, we discuss the experimental results obtained at the FEL FERMI@Elettra, and compare them with the present theoretical results. We conclude in Section~\ref{conclusions} with a brief summary.

\section{Theory}
\label{theor}

We applied the theoretical and numerical approach from the original work  \cite{PRL-Demekhin-2011} to simulate the final population of singly-ionized neon dimers and neon monomers in the process~(\ref{Eq1}). It is based on a general formulation of the time-dependent theory for the nuclear wave packet propagation (see, e.g., Refs.~\cite{JCP-Cederbaum-1993ab,JCP-Pahl-1996} and references therein), which has recently been extended  to evaluate the  resonant Auger decay  of diatomic  molecules in intense laser fields~\cite{PRL-Cederbaum-2011,PRA-Demekhin-2011b,JPB-Demekhin-2013a},  to study  ionization via multiple excitation of dimers~\cite{JPB-Demekhin-2013b}, as well as to investigate  light-induced conical intersections in polyatomic molecules~\cite{JCP-Demekhin-2013}. The extended approach and all necessary derivations of the theory can be found in Refs.~\cite{JCP-Pahl-1996,PRA-Demekhin-2011b,JPB-Demekhin-2013a,JCP-Demekhin-2013,PRA-Demekhin-2011a}. Equations describing the presently studied process are listed and discussed in the supplemental material of the original work Ref.~\cite{PRL-Demekhin-2011}. Therefore, only  essential points of the theory are outlined below.

We solve the time-dependent Schr\"{o}dinger equation to describe the interaction of Ne$_2$ with the laser pulse. To this end, we expand the total wave function of the system in terms of the relevant field-free stationary electronic states being `dressed' by the energy of photons which were absorbed in order to access these states. The presently implied \emph{Ansatz} includes the ground state of the Ne$_2(^1\Sigma_{g}^+)$,  intermediate   OV-ionized Ne$_2^+(2p^{-1})$  states $^2\Pi_{g/u}$ and $^2\Sigma_{g/u}^+$, and decaying IV-ionized Ne$_2^+(2s^{-1})$ states $^2\Sigma_{g/u}^+$ with the outgoing  photoelectron (see Fig.~\ref{Fig1}b). In the local approximation~\cite{JPB-Cederbaum-1981,PR-Domcke-1991}, the time evolution of the final $\mathrm{Ne}^+(2p^{-1})\mathrm{Ne}^+(2p^{-1})$ doubly-ionized~OVOV states can be decoupled from the dynamics of the main ansatz. Since including the doubly-ionized states is not required for the interpretation of the present experimental results (see Section~\ref{experiment}), those doubly-ionized OVOV states are not discussed below.

The time-dependent expansion coefficients in the total wave function ansatz depend explicitly on the nuclear vibrational and rotational coordinates and play a role of the two-dimensional nuclear wave packets  propagating on the potential energy surfaces of the included electronic states \cite{JCP-Cederbaum-1993ab,JCP-Pahl-1996}. The time evolution of these nuclear wave packets is governed by an effective Hamiltonian (Eq.~(S8) in the supplemental material document to the original work \cite{PRL-Demekhin-2011}), which was obtained in the rotating wave  and local  approximations.
In the case of only one OV-ionized state of  $^2\Pi_g$ symmetry resonantly coupled by the field with the IV-ionized state of $^2\Sigma^+_u$ symmetry it explicitly reads (extension to all participating states is straightforward):
\begin{eqnarray}
\fl
\hat{\mathbf{H}}(R,\theta,t)= \hat{\textbf{T}}(R,\theta)+  \nonumber\\
\fl ~~~~~~~~~~~~\left(\begin{array}{l|l|l} V_I(R)-\frac{i}{2}\Gamma_I^{ph}(t)~ &0&0\\ \hline
d_x(t)\sin \theta +  &V_{OV}(R)-\frac{i}{2}\Gamma_{OV}^{ph}(t)+  & \left(D^\dag_x(t)-\frac{i}{2} W^\dag(t)\right) \sin \theta \\
+d_z(t)\cos \theta &+\varepsilon_{ph}-\omega\ & \\ \hline
0&\left(D_x(t)-\frac{i}{2} W(t)\right) \sin \theta~  & V_{IV}(R) +\varepsilon_{ph}-2\omega -\\
& &  -\frac{i}{2}[\Gamma^{ICD}_{IV}(R)+\Gamma_{IV}^{ph}(t)]  \\ \end{array} \right), \label{Eq2}
\end{eqnarray}
where $\hat{\textbf{T}}$  is the nuclear kinetic energy operator. Let us briefly discuss this effective Hamiltonian with the emphasis on the incorporated  physical  processes evoked by intense laser pulses.

The driving pulse transfers the nuclear wave packet from the ground electronic state by its direct ionization (matrix element $d_x(t)\sin \theta+d_z(t)\cos \theta $) to all OV-ionized states of the dimer and the photoelectron is emitted (see Fig.~\ref{Fig1}b). Due to this  photoionization, the potential energy of the ground state $V_I(R)$ is augmented by the time-dependent imaginary term, $-\frac{i}{2}\Gamma_I^{ph}(t)$,  which describes losses of the corresponding population (i.e., by the total ionization rate~\cite{PRA-Demekhin-2011a,PRA-Liu-2010}). The potential energies of the OV- and IV-ionized states $V_{OV}(R)$ and $V_{IV}(R)$ are also augmented by the time-dependent imaginary corrections, $-\frac{i}{2}\Gamma_{OV}^{ph}(t)$ and $-\frac{i}{2}\Gamma_{IV}^{ph}(t)$ , respectively. Those corrections describe leakages of the corresponding populations  due to direct ionizations of the neighboring neutral Ne atom by the absorption of  subsequent photons to produce all possible doubly-ionized OVOV or IVOV states of the dimer. The energy of the IV-state is additionally augmented by the time-independent imaginary correction $-\frac{i}{2}\Gamma_{IV}^{ICD}(R)$, which represents its relaxation via the ICD transition into the possible OVOV doubly-ionized states.

The OV- and IV-ionized states of the dimer are resonantly coupled by the strong driving pulse. The respective non-Hermitian time-dependent coupling~\cite{PRA-Demekhin-2011a} is operative only as long as the pulse is on and consists of two parts. The direct coupling, $D_x(t)\sin \theta$, which is caused by the usual  excitation -- stimulated emission process, and by the indirect imaginary coupling, $-\frac{i}{2} W(t) \sin \theta$,  which appears if the photoionization from the OV-ionized  state and  ICD transition from IV-ionized state  are simultaneously  treated~\cite{PRA-Demekhin-2011a}. The coupled `dressed' IV- and OV-ionized states exhibit intersections of the  two-dimensional potential energy surfaces in the space of  vibrational and rotational dynamical variables (known as the light-induced conical intersections~\cite{JPB-Moiseyev-2008,JPB-Sindelka-2011,JPB-Halasz-2011}). Due to the presence of the ICD width, leakages by photoionization, and non-hermitian coupling, the potential energy surfaces are complex and generally exhibit two intersecting points where  real and imaginary parts of the two electronic energies become degenerate~\cite{JCP-Feuerbacher-2004}. The  non-adiabatic effects caused by these intersections~\cite{PRL-Cederbaum-2011,PRA-Demekhin-2011b,JPB-Demekhin-2013a,JCP-Demekhin-2013} are naturally incorporated in the present calculations.

The nuclear wave packets propagating on the potential energy surfaces of the incorporated electronic states contain all information relevant for the present process. After the driving pulse has expired and ICD of the Ne$_2^+(2s^{-1})$ states has essentially completed, the corresponding nuclear wave packets on all IV-ionized states vanish. At long times, there is  a fraction of neutral dimers in the ground electronic states which have survived the pulse. Their population is given by the  norm of the corresponding nuclear wave packet. The final norms of the nuclear wave packets propagating on the bound potential energy surfaces of the OV-ionized states  provide information on the singly-ionized dimers in the electronically stable Ne$_2^+(2p^{-1})$ states. Finally, the Ne$^+$ ion yield of the process can easily be obtained using normalization condition of the total nuclear wave packet with an additional contribution from the  repulsive OV-ionized states.

The two-dimensional nuclear dynamics calculations on the coupled complex energy surfaces were carried out by the efficient Multi-Configuration Time-Dependent Hartree (MCTDH) method~\cite{CPL-Meyer-1990} and code~\cite{Computing-Worth}. In the calculations, we utilized~\emph{ab initio} potential energy curves~\cite{PRL-Demekhin-2011,JCP-Stoychev-2008} and  ICD transition rates~\cite{JCP-Averbukh-2006} for Ne$_2$. The electron transition matrix elements were obtained from the experimental photoionization cross section of the Ne atom ($\sigma_{2p}={7.8}$~Mb at  28.4~eV~\cite{VUV-book-1996}) and the experimental $2s^{-1}$ to $2p^{-1}$  radiative decay lifetime of Ne$^+$ ion ($\tau_r\sim 0.14$~ns~\cite{PRL-Lablanquie-2000}).

\begin{figure}
\centering\includegraphics[scale=0.53]{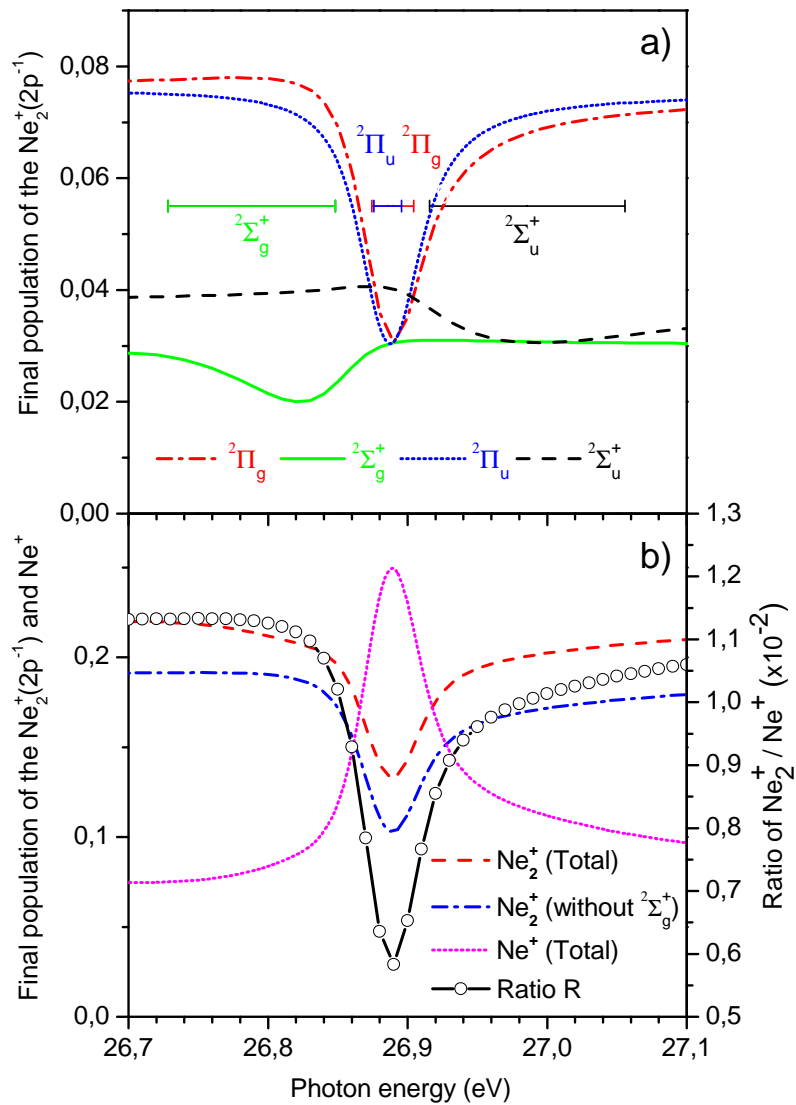}
\caption{a): Residual populations of the four OV-ionized states $\mathrm{Ne_2^{+}}(2p^{-1})$ shown in Fig.~\ref{Fig1}b after the end of the XUV pulse as functions of the photon energy. The horizontal bars give a rough estimate for the motion of the nuclear wave packet in each of the excitation pathways (see discussion in the text). b): Total residual population of all OV-ionized states $\mathrm{Ne_2^{+}}(2p^{-1})$, obtained with (red dashed line) and without (blue dash-dotted line) the contribution from the dissociative state $^2\Sigma_g^+$. The total number of the singly-ionized neon monomers generated during the excitation of dimers (see text for details) is shown by dotted curve. The ratio $R$ of the singly-ionized neon dimers $\mathrm{Ne_2^+}$ vs all singly-ionized monomers $\mathrm{Ne^+}$ is also shown as a function of the photon energy (open circles, referred to the right vertical axis). It was obtained by assuming the concentration of the neon dimers relatively to the neon atoms in the gas jet to be  $\sim 1\%$.}
\label{Fig2}
\end{figure}

\subsection{Computational results}
\label{theorRES}

Figure~\ref{Fig2}a shows the residual populations of the four OV-ionized  $\mathrm{Ne_2^+}(2p^{-1})$ states of the dimer at the end of the XUV pulse as functions of the photon energy. Calculations were performed for a Gaussian-shaped pulse of duration 70~fs FWHM of the pulse intensity envelope and peak intensity of $10^{12}$~W/cm$^2$. A clear reduction of the residual populations of the singly-ionized $\mathrm{Ne_2^+}(2p^{-1})$ dimers around the photon energy required for the resonant OV$\to$IV excitation in the ion (approximately 26.89~eV, vertical double-arrow in Fig.~\ref{Fig1}b) is visible from Fig.~\ref{Fig2}a. This is the result of the triggering of ICD, which decreases the population of $\mathrm{Ne_2^+}$ states by the ejection of a second electron and subsequent fragmentation of the neon dimer by Coulomb explosion. Each population presents a minimum for the photon energy resonant for the specific excitation pathway $\mathrm{Ne^+_2}(2p^{-1})\rightarrow \mathrm{Ne^+_2}(2s^{-1})$. Of course, nuclear dynamics accompanying the excitation step plays a role. It can be estimated by considering the FWHM of the initial nuclear wave packet in the  ground electronic state of the dimer.

The horizontal bars in Fig.~\ref{Fig2}a give a rough estimate for the motion of the nuclear wave packet in each of the excitation pathways. For this purpose, we first estimate the FWHM localization of the initial nuclear wave packet in the ground electronic state of the dimer. For each excitation pathway, the energies of the OV ionized states were then subtracted from the energies of the IV ionized states at the two internuclear distances determining the main localization of the wave packet. Thereby, the interval of localization is converted in an interval of excitation energy by assuming a vertical transition of the nuclear wave packet during the ionization and excitation steps. The estimated energy ranges are shown in Fig.~\ref{Fig2}a by the horizontal bars, and are expected to provide the main contribution to the excitation spectrum.

The minima in the populations are more pronounced for the $\Pi$ states of  $\mathrm{Ne_2^+}(2p^{-1})$ (dotted and dash-dotted curves) with respect to the $\Sigma$ states (solid and dashed curves). This is partly due to statistics (the former are doubly degenerate and the excitation probability doubles) and also due to  different underlying nuclear dynamics in the initial step for the excitation. Indeed, the potential energy curves corresponding to the $\Pi$ states of $\mathrm{Ne_2^+}(2p^{-1})$ are rather flat at the  internuclear distance which corresponds to the vertical ionization from the ground electronic state (see Fig.~\ref{Fig1}b). As a result, nuclear motion in the initial step for the excitation is very slow and plays only a moderate role. The two $\Sigma$ states of $\mathrm{Ne_2^+}(2p^{-1})$  (the uppermost and the lowest ones in Fig.~\ref{Fig1}b) are significantly steeper, and the nuclear dynamics results in the rapid propagation of the nuclear wave packet in the inward direction for the lowest state, and in the outward direction for the uppermost one. This dynamics competes with the further excitation step and makes the excitation spectrum broader.

The total residual population of all OV-ionized $\mathrm{Ne_2^+}$ states is depicted in Fig.~\ref{Fig2}b by the dashed curve. Since the $^2\Sigma_g^+$ OV-ionized state is repulsive (see Fig.~\ref{Fig1}b), it results in the formation of the $\mathrm{Ne}$ and $\mathrm{Ne^+}$ fragments. Therefore, its contribution needs to be excluded from the fraction of the singly-ionized neon dimers after the end of the pulse. The total population of the surviving singly-ionized neon dimers ($\mathrm{Ne_2^+}$ without $^2\Sigma^+_g$), corrected for the contribution of the latter state, is shown in Fig.~\ref{Fig2}b by the dash-dotted blue curve. The total number of the singly-ionized neon monomers generated during the process~(\ref{Eq1}) consists of two parts, i.e., of the ions resulting from ICD and of the ions produced by the dissociation of the $^2\Sigma_g^+$ OV-ionized  state. This population is also shown in Fig.~\ref{Fig2}b by the red dotted curve ($\mathrm{Ne^+}$ Total). It indicates that the reduction of the singly-ionized neon dimers is correlated with the increase of the singly-ionized  monomers.

We now define the ratio $R$ of the singly-ionized neon dimers vs singly-ionized neon monomers. It is very important to point out that, in typical experimental conditions, neon dimers are produced with a large background of neon monomers. Therefore, the ratio $R$ should take into account $\mathrm{Ne^+}$ produced by two mechanisms. The dominant contribution to the $\mathrm{Ne^+}$-signal stems from the single-photon direct ionization of the large fraction of the monomers. A minor contribution to the total $\mathrm{Ne^+}$-signal is provided by the  small fraction of dimers via the processes discussed in the preceding paragraph (red dotted curve $\mathrm{Ne^+}$ Total in Fig.~\ref{Fig2}b). The ratio $R$ is thus given by:
\begin{equation}
R=\frac{P(\mathrm{Ne_2^+})}{P(\mathrm{Ne^+},~\mathrm{monomers})+P(\mathrm{Ne^+},~\mathrm{dimers})}
\label{Eq3}
\end{equation}
Since the total yield of the singly-ionized neon dimers is not affected by the presence of neon monomers,  the ratio $R$ is expected to be sensitive to the excitation of ICD.

The $\mathrm{Ne^+}$ yield is expected to depend linearly on the intensity as it is produced by the direct one-photon ionization of neon monomers. Out of resonance, stable $\mathrm{Ne_2^+}$ in the OV states are produced by the one-photon ionization which presents a linear dependence on the intensity I. On resonance, part of these stable OV ionized dimers is promoted to unstable IV ionized states, which fragment by ICD. The number of surviving OV ionized dimers is expected to be proportional to $I(a-bI)$ with a and b constants. The first term $aI$ is the number of singly-ionized dimers produced by ionization, and the second term $-bI^2$ is the number of the ionized dimers promoted to the decaying states. The first term turns out to be dominant in the intensity range $5\times10^{11}-2\times10^{12}~\mathrm{W/cm^2}$, while the quadratic term $-bI^2$ determines only a small correction leading to a slower increase of the final population of $\mathrm{Ne_2^+}$ for increasing intensities.

In calculating $R$, we assumed the relative population of neon dimers in the jet to be $\sim 1\%$ (see experimental details in Section~\ref{experiment}). For small variations of the dimer concentration, the ratio $R$ scales as the ratio of the neon dimer versus the neon monomer concentration. A clear signature of ICD can be identified in this ratio $R$, which is shown in Fig.~\ref{Fig2}b by open circles as a function of the photon energy. According to the calculations,  this ratio presents a clear dip around the photon energy of about 26.89~eV resonant to the triggering of ICD.

\section{Experiment}
\label{experiment}

\begin{figure}
\centering\includegraphics[scale=0.6]{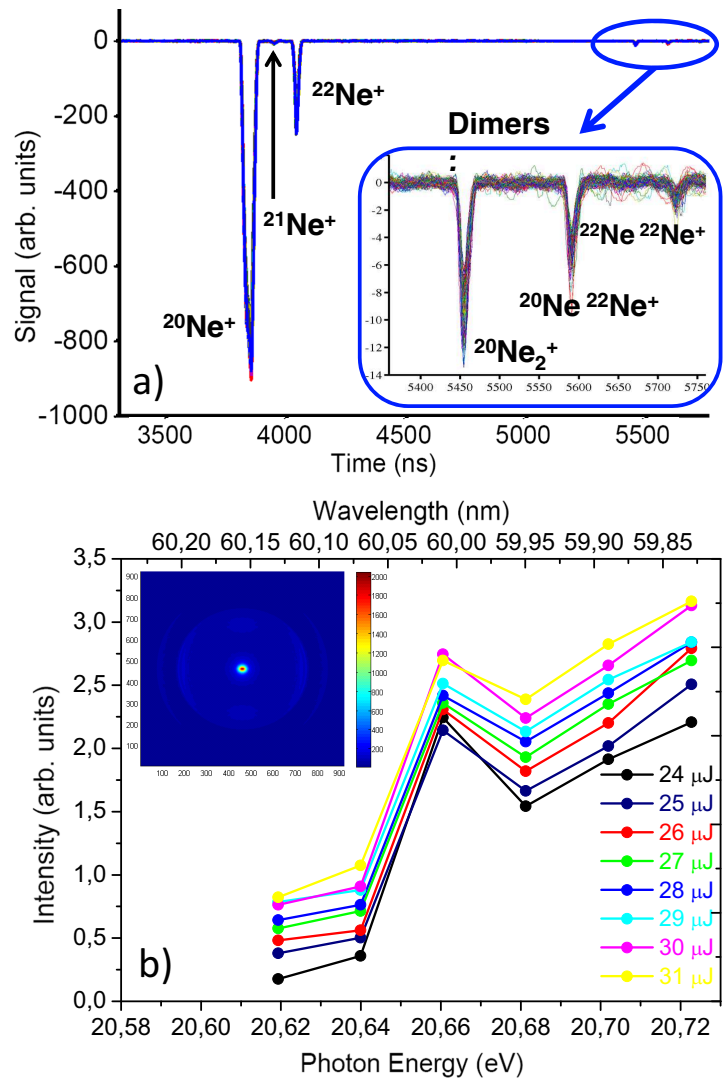}
\caption{Experimental evidence of neon dimers. a): Neon mass ion spectra. b): Photoelectron yield as a function of the photon energy for the zero kinetic energy electron peak extracted from the Velocity map imaging spectrometer (VMI) measurements (shown in the inset of the panel).} \label{Fig3}
\end{figure}

The experiment was performed on the low-density matter (LDM) end station~\cite{JPB-Lyamayev-2013} installed on the seeded FEL FERMI@ELETTRA~\cite{NATPHOT-Allaria-2012}. The seeding process ensures the optimal shot-to-shot spectral and energy stability required for the investigation of resonant nonlinear processes in the XUV regime. The pulse energy was monitored on a single-shot basis by two ionization monitors placed just before and after a gas attenuation cell, which was used to finely adjust the pulse energy. The measurement of the second cell was used for the post-processing of the experimental data~\cite{RSI-Zangrando-2009}.

After the on-line monitors, the XUV pulses were directed towards the end station by a plane grating, which was weakly ruled over a part of its surface so that it reflected most of the incident radiation, while a small fraction of about $1\%$ was diffracted in the first order and acquired by a CCD camera on a single-shot basis. The XUV pulses were then reflected by a plane mirror and focused by a Kirkpatrick-Baez~(KB) arrangement. The total throughput of the XUV photon transport beam line was estimated to be $45\%$. The focal spot was optimized in the interaction chamber using a movable Yag screen. A radius spot size of about 60~$\mathrm{\mu m}$ was measured under typical operating conditions. We estimated an intensity of $1.4\times10^{12}~\mathrm{W/cm^2}$, considering a pulse duration of~70~fs and a pulse energy of 26~$\mathrm{\mu J}$.

Neon dimers were generated using an Even-Lavie valve at room temperature and at a backing pressure of~16~bar. The concentration of neon dimers was optimized using a quadrupole mass spectrometer installed in the experimental end station after the interaction region. An optimal concentration of about $1\%$ of dimers with respect to monomers was achieved by fine tuning of the backing pressure. For higher backing pressure larger clusters were also observed in the molecular jet.

Ion and electron measurements were performed using the detectors installed on the LDM end station. Figure~\ref{Fig3}a shows the ion mass spectra measured using XUV pulses centered at the photon energy $\hbar\omega=26.9$~eV. The spectrum is dominated by singly-ionized neon monomers $\mathrm{^{20}Ne^+}$ and $\mathrm{^{22}Ne^+}$, with a small fraction of neon dimers (see inset). The concentration of neon dimers considering the integral of the monomer and dimer peaks was consistent with the measurement of the quadruple mass spectrometer.

In the experiment we have verified that both $\mathrm{Ne^+}$ and $\mathrm{Ne_2^+}$ yields present a dominant linear dependence in the investigated range of pulse intensities, in agreement with the simulations.

The presence of neon dimers in the molecular beam was confirmed by measuring the photoelectron yield as a function of the photon energy below the ionization threshold of neon atoms. The inset of Fig.~\ref{Fig3}b shows the photoelectron position spectrum measured at 20.59~eV photon energy. The spectrum is dominated by a strong central peak of photoelectrons with almost zero kinetic energy. The integral of this peak as a function of the photon energy is shown in Fig.~\ref{Fig3}b. The sharp increase at 20.62~eV and the non-monotonic evolution indicates that the central peak originates from single-photon ionization of neon dimers~\cite{JCP-Trevor-1984}. Indeed in this energy range single-photon ionization of neon monomers is not energetically allowed and the structure in the yield around 20.66--20.88~eV is due to an autoionization feature in the neon dimer cross section~\cite{JCP-Trevor-1984}. A partial contribution of larger clusters to the low energy electron spectrum cannot be completely ruled out.

\begin{figure}
\centering\includegraphics[scale=0.58]{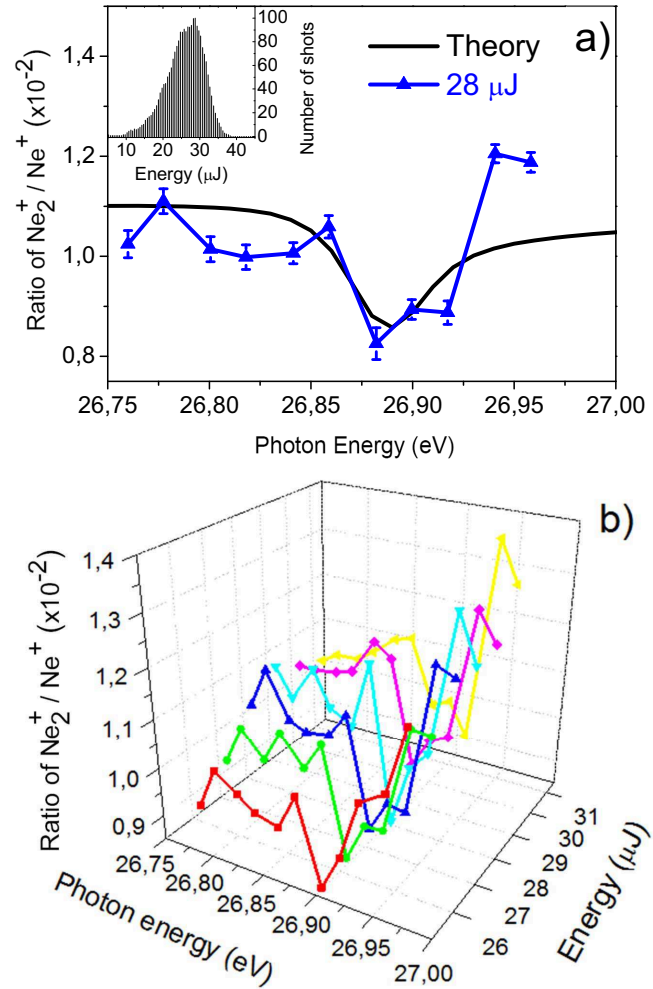}
\caption{a): Ratio \emph{R} as a function of the photon energy measured for a fixed pulse energy (symbols). The error bar is defined as the standard deviation of the mean. The solid curve represents the present theoretical result, which was additionally convolved with photon energy bandwidth of 50~meV FWHM. A typical single-shot pulse energy distribution acquired for a fixed photon energy is shown in the inset. b): Ratio \emph{R} as a function of the photon energy for different XUV pulse energies corresponding to the center of the distribution shown in the inset of panel a). c): Ratio \emph{R} as a function of the pulse energy for the on (filled squares) and out of resonance (open circles and diamonds) cases.}
\label{Fig4}
\end{figure}

As suggested by the calculations reported in section~\ref{theorRES}, we have measured the ion mass spectra as a function of the photon energy in the range of 26.75--26.95~eV in steps of 20~meV. For each central photon energy, we simultaneously acquired the single-shot XUV energy. The inset of Fig.~\ref{Fig4}a shows the histogram of the XUV pulse energies for a fixed photon energy. By varying the photon energy, small variations in the XUV pulse energy occurred, which were compensated for by tuning the pressure in the gas attenuator. The pulse energy was adjusted in the range of 25--30~$\mu J$.  For each photon energy, the shots were divided in 1~$\mathrm{\mu J}$ wide energy bins and only bins with more than 25 measurements were retained in the data analysis. For each shot we calculated the ratio $R$ between the integral of the peak of the neon monomer isotope $\mathrm{^{22}Ne^+}$ and the integral of the neon dimer peak $\mathrm{^{20}Ne_2^+}$. The integral of this isotope of $\mathrm{Ne^+}$ was preferred in order to avoid any artefact due to saturation of the detector, which occasionally occurred due to the high signal of the $\mathrm{^{20}Ne^+}$ peak. The ratio was then normalized for the relative natural abundance of the two neon isotopes.

Figure~\ref{Fig4}a reports the measured ratio $R$  obtained for the XUV pulse energy of 28~$\mathrm{\mu J}$ (solid triangles). The signal presents a clear dip at the photon energy resonant to the OV$\to$IV excitation in the Ne$_2^+$ ion (around 26.89~eV), and matches well with the theoretical expectations (solid curve). Good agreement between the theoretical and experimental ratios $R$ confirms the possibility to trigger ICD via the absorption of two photons in process (\ref{Eq1}). The error bars in Fig.~\ref{Fig4}a represent the standard deviation of the single shot distribution for each experimental point of the curve. The data have been sorted and filtered out according to the single shot pulse energy and spectrum. Therefore experimentally, the main contribution to the results dispersion is due the intensity fluctuations at focus (due to pulse-to-pulse fluctuations of the temporal duration and beam profile). However, according to the error bars, those fluctuations are sufficiently small to allow the dip in the ratio to be clearly identified. The present observation of the two-photon excitation of ICD is also robust against variations of the XUV pulse energy as demonstrated in Fig.~\ref{Fig4}b. It depicts the ratios  $R$ as a function of the photon energy acquired at different energies of the pulse (see the histogram  in the inset to Fig.~\ref{Fig4}a). Finally Fig.~\ref{Fig4}c reports the evolution on and out of resonance of the ratio $R$ for different pulse energies.

\section{Conclusions}
\label{conclusions}

A combined theoretical and experimental study of the two-photon excitation of ICD by intense FEL pulses is reported. By measuring the ion mass spectra as functions of the central photon energy and analyzing the distribution of singly-ionized dimers and monomers we demonstrate the possibility to initiate ICD in neon dimers by the absorption of two photons. The experimental data are in agreement with the \emph{ab initio} theoretical results, both showing a sharp decrease of the surviving singly-ionized neon dimer population when ICD occurs. The experiments took full advantage of the unique characteristics (energy tunability, narrow bandwidth, and high intensity) of the XUV pulses delivered by  FERMI@Elettra. The present results open new perspectives for the investigation of the electron-correlation-driven relaxation mechanisms in clusters by multiphoton absorption schemes.

\ack
Financial support by the Alexander von Humboldt Foundation (Project `Tirinto'), the Italian Ministry of Research (Project FIRB  No. RBID08CRXK), the European Research Council under the European Community's Seventh Framework Programme (FP7/2007-2013) / ERC grant agreements no.~227355 - ELYCHE and 227597 - ICD, the State Hessen Initiative LOEWE within the focus project ELCH is gratefully acknowledged. KM and KU are grateful for support by the x-ray Free Electron Laser Priority Strategy Program of MEXT and by Tagen project by IMRAM. SM is grateful to JSPS for support. This project has received also funding from the European Union's Horizon 2020 research and innovation programme under the Marie Sklodowska-Curie grant agreement no.~641789 MEDEA.

\end{document}